
  \def\sa{\vskip 0.5 true cm} 
  \def\sb{\vskip 1   true cm}
  \def\avant{\vskip 0.4 true cm}
  \def\apres{\vskip 0.2 true cm}

\magnification\magstep1
\vsize = 22 true cm
\hsize = 16.8 true cm
\parskip = 0.50 true cm

\pageno = 0


   \hfill \vbox{\hsize 2.8 true cm 
   \noindent
   \bf LYCEN 9307\break 
   May 1993}

\medskip

\baselineskip = 0.65 true cm
 
  \sb
  \sa

\noindent{\bf Sum Rules for Multi-Photon 
Spectroscopy of Ions in Finite Symmetry}

\sb
\sa

\noindent{M.~KIBLER and M.~DAOUD} 

\sa

{\it 
\noindent{Institut de Physique Nucl\'eaire, 
IN2P3~$-$~CNRS et Universit\'e Lyon-1,}

\noindent{43 Boulevard du 11 Novembre 1918, 
69622 Villeurbanne Cedex, France}
}

\sb
\sb
\sb

\noindent 
{\bf  Abstract}. 
Models describing one-  and two-photon transitions for ions in 
crystalline environments  are unified and extended to the case 
of parity-allowed and parity-forbidden $p$-photon transitions. 
The number of independent parameters  for  characterizing  the 
polarization dependence  is shown to  depend on an ensemble of
properties and rules which combine symmetry considerations and 
physical models. 

\sb


\sb

\noindent Published in 
          {\bf Letters in Mathematical Physics 28 (1993) 269-280}. 

\vfill\eject
\baselineskip = 0.90 true cm 

\noindent {\bf 0. Introduction and Preliminaries} 

\apres

\noindent Electronic spectroscopy of partly-filled shell ions, 
like $d^N$ and $f^N$ ions, embedded in condensed matter 
surroundings has been the object of important developments in 
the 
recent years. In particular, the advent of tunable dye lasers 
has made possible to perform two-photon spectroscopy experiments. 
This Letter is concerned with $p$-photon 
spectroscopy of a transition ion with ground atomic configuration 
$n \ell^N$ in a given crystal field with generic symmetry 
$G$. We address here the question of obtaining a closed-form 
expression for the intensity of a $p$-photon 
(absorption) transition between 
an initial Stark level $i$ with symmetry species $\Gamma$ and a 
     final Stark level $f$ with symmetry species $\Gamma'$. The 
labels $\Gamma$ and $\Gamma'$ stand for two IRC's (irreducible 
representation classes) of the double group $G^*$ of 
$G \simeq G^*/Z_2$. The group $G^*$ may be considered (up to 
isomorphism) as a subgroup of $SU(2)$. 

According to Dirac and G\"oppert-Mayer, the transition 
moment $M_{i(\Gamma \gamma) \to f(\Gamma' \gamma')}$ 
for a $p$-photon transition between the initial state 
$i(\Gamma \gamma)$ and the final state $f(\Gamma' \gamma')$ 
is given by 
$$
M_{i(\Gamma \gamma) \to f(\Gamma' \gamma')} = \sum_{ v_j }
{ \left(f \Gamma' \gamma'       
                \vert ( {\bf D} . \, {\bf E}_1 ) \vert v_1 \right) 
  \left(v_1     \vert ( {\bf D} . \, {\bf E}_2 ) \vert v_2 \right)
  \cdots 
  \left(v_{p-1} \vert ( {\bf D} . \, {\bf E}_p ) \vert i \Gamma \gamma \right) 
\over 
(\Omega_i - \Omega_{v_1    }) 
(\Omega_i - \Omega_{v_2    }) \cdots
(\Omega_i - \Omega_{v_{p-1}}) } + {\rm perm} 
\eqno (1)
$$
in the framework of the electric dipolar approximation. 
The sum on $v_j$ in (1) has to be performed on 
all the intermediate states and ``perm'' indicates that 
$p!$ permuted terms arise in the Feynman representation associated 
to $M_{i(\Gamma \gamma) \to f(\Gamma' \gamma')}$ 
when the $p$ photons are different. The operator 
$( {\bf D} . \, {\bf E}_k )$ 
in (1) is the scalar product of the 
dipole moment operator ${\bf D}  $ for the $N$ electrons with the unit 
polarization vector    ${\bf E}_k$ of the single-mode for the $k$th 
photon. (We use single-modes to describe 
the electromagnetic field.) Furthermore, the factors 
$(\Omega_i - \Omega_{v_{j}})$ 
are energy denominators 
for the system radiation field plus ion in its environment. 
Finally, the labels 
$\gamma$ and $\gamma'$ distinguish the components of the Stark levels 
$i$ and $f$ when the dimensions $[\Gamma]$ and $[\Gamma']$ 
of the IRC's $\Gamma$ and $\Gamma'$ are greater than 1, respectively. 

At this stage, we must consider two types of $p$-photon 
transitions: the parity-allowed and parity-forbidden transitions. 
The parity-allowed transitions occur either between levels of the 
same parity ($i \in n \ell^N$ and $f \in n \ell^N$) 
when $p$ is even or between levels of opposite parities 
($i \in n \ell^N$ and $f \in n \ell^{N-1} n' \ell'$ with 
$(-1)^{\ell + \ell'} = -1$) when $p$ is odd. 
The parity-forbidden transitions 
                         ($n \ell^N \to n \ell^N$ for $p$ odd and 
                          $n \ell^N \to n \ell^{N-1} n' \ell'$ with 
$(-1)^{\ell + \ell'} = -1$ 
for $p$ even) may become possible, via the Van Vleck mechanism, 
if $G$ does not have a center of inversion (or if the inversion 
symmetry is broken by vibronic coupling or ligand polarization). 

The sums over the intermediate states $v_j$ in (1) can be 
effectuated by using a quasi-closure approximation 
of the type of the ones used by many authors [1-5] 
for $p=1$ or $2$ in the cases of $d^N$ and $f^N$ ions. Such 
an approximation used in conjunction with recoupling 
techniques for $SU(2)$ irreducible tensorial sets leads to 
$$
M_{ i(\Gamma \gamma) \to f(\Gamma' \gamma') } 
= ( f \Gamma' \gamma' \vert H_{\rm eff} \vert i \Gamma \gamma ), 
\eqno (2)
$$
where the model-dependent operator $H_{\rm eff}$ is 
an effective operator described in Section 2.

Both for parity-allowed 
and      parity-forbidden transitions, we take 
the initial and final state vectors in the symmetry adapted form 
($x \equiv i$ or $f$) 
$$
\vert x \Gamma \gamma ) \; = \; 
\sum_{\alpha S L J a} \; \vert \alpha S L J a \Gamma \gamma) 
 \; c(\alpha S L J a \Gamma; x). 
\eqno (3)
$$
The initial state $i(\Gamma  \gamma )$ belongs 
to the configuration $n \ell^{N}$ 
and   the final   state $f(\Gamma' \gamma')$ belongs either 
to the configuration $n \ell^{N}$ or 
to the configuration $n \ell^{N-1} n' \ell'$. 
In (3), the $SU(2) \supset G^*$ vectors of type 
$\vert j a \Gamma \gamma)$ 
are connected to the $SU(2) \supset U(1)$ 
vectors of type $\vert jm)$ by 
$$
\vert j a \Gamma \gamma) \; := \; \sum_{m=-j}^j \; 
\vert j m ) \; (j m \vert j a \Gamma \gamma). 
\eqno (4)
$$
The branching label $a$ in (4) has to be used when the IRC $\Gamma$ of 
$G^*$ occurs several times in the IRC $(j)$ of $SU(2)$. This 
external multiplicity label may be described, at least 
partially, by IRC's of subgroups of $SU(2)$ that contain in 
turn the group $G^*$. The expansions (3) and (4) are of 
a completely different nature. Equation (4) is of a 
group theoretical nature: the reduction coefficients 
$(jm \vert j a \Gamma \gamma)$ depend on the chain $SU(2) \supset G^*$ 
but not on the physics of the problem. On the other hand, the 
expansions (3) depend on physical models: the $c$ coefficients 
may be obtained from the diagonalization of one or two $G$-invariant 
Hamiltonians (according to whether as the initial and final 
configurations are the same or different) describing the ion in 
its environment. 
It is to be noted that, thanks to Schur's lemma, we 
may  always  standardize the reduction coefficients 
    $(j m \vert j a \Gamma \gamma)$ in (4) in such a way 
that the expansion coefficients $c$ in (3) be independent of the 
generalized magnetic quantum numbers $\gamma$ and $\gamma'$. 
This result is essential in view of the sum on $\gamma$ and $\gamma'$ 
in (5) below. Equation (3) indicates that 
we adopt here the philosophy of [6-9] 
where the diagonalization-optimization of the $G$-invariant 
Hamiltonian(s) is done, without loss of generality, 
in a weak-field 
basis adapted to the chain $SU(2) \supset G^*$: the labels $\alpha$, $S$, 
$L$ and $J$ are not taken {\it a priori} as good quantum numbers; the 
only good quantum numbers are $\Gamma, \gamma$ and $\Gamma', \gamma'$ for 
the initial and final state vectors, respectively 
(and the parity for parity-allowed transitions). 

The physical quantity we want to calculate 
in  Section  2  is  the  intensity
$$
S_{ i(\Gamma) \to f(\Gamma') } \; := \; \sum_{\gamma \gamma'} \; 
\left\vert M_{i(\Gamma \gamma) \to f(\Gamma' \gamma')} \right\vert ^2 
\eqno (5)
$$
for the $i(\Gamma) \to f(\Gamma')$ transition. 
The sum in (5) has to be done 
over all the external multiplicity labels $\gamma$ and 
$\gamma'$, i.e., over all the components of the initial and 
final Stark levels. (The labels $\gamma$ and $\gamma'$
may be characterized, at least 
partially, by IRC's of subgroups of $G^*$.) 
The purpose of this work is to show how far group theory can go 
in the determination of $S_{i(\Gamma) \to f(\Gamma')}$ 
in connection with sophisticated models for the operator 
$H_{\rm eff}$ of (2). In particular, we want to clearly 
separate the elements arising from group theory (via 
Wigner-Racah calculus) from the model-dependent quantities 
(both for the light-matter interaction and   
the description of the $\ell^N$ ion in its environment). 
Rather than considering $G^*$ as an isolated group, we shall 
consider it as a subgroup of $SU(2)$, an approach that has 
already proved to be very useful as far as the energy spectrum 
of the $\ell^N$ ion in $G$ symmetry is concerned [8,~9]. 
The basic tool is thus the Wigner-Racah algebra of the group $G^*$ 
in a $G^* \subset SU(2)$ basis. Therefore 
we develop, in Section 1, 
those nontrivial aspects of the Wigner-Racah 
algebra for the chain $SU(2) \supset G^*$ that are essential 
for performing the sum in (5). The 
corollary obtained in Section 1 is applied in Section 2 to 
derive the sum rule (23) 
with the accompanying properties 
and rules (26-28). 

\avant

\noindent {\bf 1. Sum Rule for $SU(2) \supset G^*$ Coupling Coefficients}

\apres

\noindent Equation (4) defines the $\left\{ j a \Gamma \gamma \right\}$ 
scheme for the chain $SU(2) \supset G^*$ that is more 
appropriate in condensed matter spectroscopy than the 
$\left\{ j m \right\}$ scheme for the chain $SU(2) \supset U(1)$ 
[6,~8,~9]. For $j$, $a$ and $\Gamma$ fixed, the set 
$\left\{ \vert j a \Gamma \gamma) \ : \ \gamma = 1, \cdots, [\Gamma] \right\}$ 
is a (standardized) $G$-irreducible tensorial set (in the sense of 
Fano and Racah) of vectors associated to $\Gamma$. 
Similarly, from the $SU(2) \supset U(1)$ spherical tensor 
operators $U^{(k)}_q$, we define the operators 
$U^{(k)}_{a \Gamma \gamma}$ by an expansion similar to (4) 
so that, for $k$, $a$ and $\Gamma$ fixed, the set 
$\left\{ U^{(k)}_{a \Gamma \gamma} \ : \ \gamma = 1, \cdots, [\Gamma] \right\}$ 
is a $G$-irreducible tensorial set of operators associated to $\Gamma$.
The latter $G$-irreducible tensorial sets are also labelled by 
IRC's of $SU(2)$ and, therefore, we can easily generate, by direct 
sum, nonstandard $SU(2)$-irreducible tensorial sets. Thus, we may 
apply the Wigner-Eckart theorem for the group $SU(2)$ in a 
nonstandard basis adapted to its subgroup $G^*$. As a result, 
we have [8] 
$$
\eqalign{
  (\tau_1 j_1 a_1 \Gamma_1 \gamma_1 \vert U^{(k)}_{a \Gamma \gamma} \vert 
   \tau_2 j_2 a_2 \Gamma_2 \gamma_2) \; = \;
  (\tau_1 j_1 \Vert U^{(k)} \Vert 
   \tau_2 j_2) & \cr 
  \sum_{a_3 \Gamma_3 \gamma_3}
  \pmatrix{
                        & j_1 &                        \cr\cr
  a_3 \Gamma_3 \gamma_3 &     & a_1 \Gamma_1 \gamma_1} & \ \; 
  \overline f
  \pmatrix{
  j_1                   & k               & j_2 \cr\cr
  a_3 \Gamma_3 \gamma_3 & a \Gamma \gamma & a_2 \Gamma_2 \gamma_2}^*
  }
  \eqno (6)
$$
with 
$$
\eqalign{
   \overline f 
   \pmatrix{
   j_1                   & j_2                   & j_3\cr\cr
   a_1 \Gamma_1 \gamma_1 & a_2 \Gamma_2 \gamma_2 & a_3 \Gamma_3 \gamma_3} 
   \ := \ \sum _{m_1 m_2 m_3} \ 
  &\pmatrix{
   j_1 & j_2 & j_3
   \cr\cr
   m_1 & m_2 & m_3}
   \cr 
   (j_1 m_1   \vert j_1 a_1 \Gamma_1 \gamma_1 )^* \ 
  &(j_2 m_2   \vert j_2 a_2 \Gamma_2 \gamma_2 )^* \
   (j_3 m_3   \vert j_3 a_3 \Gamma_3 \gamma_3 )^* 
   }
   \eqno (7)
$$
and
$$
  \pmatrix{
                        & j &                         \cr\cr
  a_1 \Gamma_1 \gamma_1 &   &a_2 \Gamma_2 \gamma_2\cr}
  \ := \ 
  [j]^{1 / 2} \ \; 
  \overline f 
  \pmatrix{
  j                     & 0                     & j\cr\cr
  a_1 \Gamma_1 \gamma_1 & a_0 \Gamma_0 \gamma_0 & a_2 \Gamma_2 \gamma_2},
  \eqno (8) 
$$
where $\Gamma_0$ denotes the identity IRC of $G^*$.   In 
the right hand side of (6), the quantum numbers $\tau_1$ 
and $\tau_2$, external to the chain $SU(2) \supset G^*$, 
appear in the reduced matrix element. 

The $\overline f$ or $3$-$j a \Gamma \gamma$ symbol in (7) is an 
$SU(2) \supset G^*$ symmetry adapted form of the usual $3$-$jm$ 
Wigner symbol. It constitutes a symmetrized form of the coefficient 
[8]
$$
  \eqalign{
  f 
  \pmatrix{
  j_1                   & j_2                   & k\cr\cr
  a_1 \Gamma_1 \gamma_1 & a_2 \Gamma_2 \gamma_2 & a \Gamma \gamma}
  \ := \ & \sum_{m_1 q m_2} \;
  (- 1)^{j_1 - m_1} \;
  \pmatrix{
  j_1  & k & j_2 
  \cr\cr
  -m_1 & q & m_2}
  \cr 
& (j_1 m_1 \vert j_1 a_1 \Gamma_1 \gamma_1)^* \
  (k   q   \vert k   a   \Gamma   \gamma)     \
  (j_2 m_2 \vert j_2 a_2 \Gamma_2 \gamma_2)   \cr
  }
  \eqno (9)
$$
that generalizes the $f$ coefficient defined 
by Racah and some of his students (in particular Sch\"onfeld 
and Flato, see references [6,~7]). 

The $2$-$j a \Gamma \gamma$ symbol defined by (8) 
turns out to be an $SU(2) \supset G^*$ symmetry 
adapted form of the Herring-Wigner 
metric tensor. 
Indeed, from (7) and (8) we have
$$
  \pmatrix{
                        & j &                       \cr\cr
  a_1 \Gamma_1 \gamma_1 &   & a_2 \Gamma_2 \gamma_2}
  \ = \ \sum_{m} \ (-1)^{j + m} \
  (j  m \vert j a_1 \Gamma_1 \gamma_1)^* \
  (j,-m \vert j a_2 \Gamma_2 \gamma_2)^*. 
  \eqno (10)
$$
Therefore, the metric tensor given by (10) allows us to handle all 
the phases occurring in the $\left\{ j a \Gamma \gamma \right\}$ 
scheme [8]. 

By combining (7), (9) and (10), we can rewrite (6) in the 
simple form
$$
  (\tau_1 j_1 a_1 \Gamma_1 \gamma_1 \vert U^{(k)}_{a \Gamma \gamma} \vert 
   \tau_2 j_2 a_2 \Gamma_2 \gamma_2) \ = \ 
  (\tau_1 j_1 \Vert U^{(k)} \Vert \tau_2 j_2) \ \; 
  f \pmatrix{
  j_1                   & j_2                   & k\cr\cr
  a_1 \Gamma_1 \gamma_1 & a_2 \Gamma_2 \gamma_2 & a \Gamma \gamma}. 
  \eqno (11)
$$
The interest of (6) and (11) for electronic spectroscopy of ions in 
crystalline fields has 
been discussed in [8,~9]. From a mathematical 
viewpoint, it is to be observed that the factorization in 
(11) into the product of a $G^*$-dependent 
factor (the $f$ coefficient) 
by a $G^*$-independent factor is valid whatsoever 
the group $G^*$ is multiplicity free or not. The internal 
multiplicity problem arising when $G^*$ is not 
multiplicity free is thus easily solved by making use of (11). 

The $f$ coefficients may be considered as the components of 
a third-rank tensor (second-rank covariant and first-rank 
contravariant). In contradistinction, the ${\overline f}$ 
symbol defines a third-rank contravariant tensor. The other 
properties of the ${\overline f}$ and $f$ coefficients that 
are of relevance for this work are based on the following lemma. 

\noindent LEMMA (Racah's lemma). 
{\it We have the factorization formula [10] 
$$
  \eqalign{
  f \pmatrix{
  j_1 & j_2 & k\cr
  \cr
  a_1 \Gamma_1 \gamma_1&a_2 \Gamma_2 \gamma_2&a \Gamma \gamma} =&
  \; (-1)^{2k} \; [j_1]^{- 1/2}\cr
& \sum_{\beta} \ 
  (j_2 a_2 \Gamma_2 + k a \Gamma \vert j_1 a_1 \beta \Gamma_1)^* \ 
  (\Gamma_2 \Gamma \gamma_2 \gamma \vert 
   \Gamma_2 \Gamma \beta \Gamma_1 \gamma_1)^*, 
  } 
\eqno (12)
$$
where 
$(j_2 a_2 \Gamma_2 + k a \Gamma \vert j_1 a_1 \beta \Gamma_1)$ 
is an isoscalar factor or {\it reduced} $f$ {\it symbol} 
(independent of the quantum numbers $\gamma_1$, $\gamma_2$ and $\gamma$) and 
$(\Gamma_2 \Gamma \gamma_2 \gamma \vert 
  \Gamma_2 \Gamma \beta \Gamma_1 \gamma_1)$ 
a Clebsch-Gordan coefficient for the group $G^*$ considered as 
an isolated entity. (The label $\beta$ is an internal 
multiplicity label to be used when the Kronecker product 
$\Gamma_2 \otimes \Gamma$ is not multiplicity free.) 
}

{\it Proof}. Equation (12) is a consequence of Schur's lemma. 

In order to be able to use tables and/or computer 
     programs for the highly symmetrical ${\overline f}$ symbol 
(and for the $R_{\beta}$ or {\it reduced ${\overline f}$ symbol}, 
see (13) and (14)), 
it is useful to transcribe (12) for 
the ${\overline f}$ symbol. We thus obtain 
$$
  \overline f
  \pmatrix{
  j_1                   & j_2                   & j_3\cr\cr
  a_1 \Gamma_1 \gamma_1 & a_2 \Gamma_2 \gamma_2 & a_3 \Gamma_3 \gamma_3}
  \; = \; \sum_{\beta} \ 
  R_{\beta} 
  \pmatrix{
  j_1          & j_2          & j_3\cr\cr
  a_1 \Gamma_1 & a_2 \Gamma_2 & a_3 \Gamma_3} \ \, 
  V_{\beta} 
  \pmatrix{
  \Gamma_1 & \Gamma_2 & \Gamma_3\cr\cr
  \gamma_1 & \gamma_2 & \gamma_3}, 
  \eqno (13)
$$
where $V_{\beta}$ is an extension, taking into account 
the multiplicity label ${\beta}$, of the $V$ coefficient introduced 
by Griffith [11]. 
The coefficient 
$$
  R_{\beta} 
  \pmatrix{
  j_1          & j_2          & j_3\cr
  \cr
  a_1 \Gamma_1 & a_2 \Gamma_2 & a_3 \Gamma_3} 
  \equiv 
  R_{\beta} 
  \pmatrix{
  j_1          & j_2          & j_3\cr
  \cr
  a_1 \Gamma_1 & a_2 \Gamma_2 & a_3 \Gamma_3}^{SU(2)}_{G^*}
  \eqno (14)
$$
characterizes the restriction 
from $SU(2)$ to $G^*$; it is a generalization of the $V$ symbol 
introduced by the Chinese school of Tang Au-chin [12]. (In (13) 
and (14), the label $\beta$ is absolutely necessary when the 
identity IRC 
$\Gamma_0$ appears several times in 
$\Gamma_1 \otimes
 \Gamma_2 \otimes
 \Gamma_3$.) 
The $R_{\beta}$ symbol is connected to the isoscalar 
factor of (12) by 
$$
  \eqalign{
  (j_2 a_2 \Gamma_2 + k a \Gamma \vert j_1 a_1 \beta_1 \Gamma_1) = 
  \; & (- 1)^{2k} \; 
  [\Gamma_1]^{-1} \; [j_1] \; \sum_{a_3 \Gamma_3} \; \sum_{\beta_2} 
                  \; U(\Gamma_1 \Gamma_2 \Gamma)_{\beta_1 \beta_2} \cr 
& 
  R_{\beta_2}
  \pmatrix{
  k       & j_2         & j_1\cr
  \cr
  a\Gamma & a_2\Gamma_2 & a_3\Gamma_3} \;
  R
  \pmatrix{
  0            & j_1          & j_1\cr
  \cr
  a_0 \Gamma_0 & a_1 \Gamma_1 & a_3 \Gamma_3}^*, \cr
  } 
  \eqno (15)
$$
where $U(\Gamma_1 \Gamma_2 \Gamma)$ is an 
arbitrary unitary matrix. Indeed, this 
matrix is a simple phase factor when the Kronecker product 
$\Gamma_2 \otimes \Gamma$ is multiplicity free. 

Repeated applications of Racah's lemma and of the 
orthonormality-completeness property for the Clebsch-Gordan 
coefficients of $G^*$ lead to the following corollary. 

\noindent COROLLARY. 
{\it The sum
$$
  X :=
  \sum_{\gamma_1 \gamma_2} \; f 
  \pmatrix{
  j_1                   & j_2                   & k_3\cr
  \cr
  a_1 \Gamma_1 \gamma_1 & a_2 \Gamma_2 \gamma_2 & a_3 \Gamma_3 \gamma_3}^*\;
  f
  \pmatrix{
  j_4                   & j_5                   & k_6\cr
  \cr
  a_4 \Gamma_1 \gamma_1 & a_5 \Gamma_2 \gamma_2 & a_6 \Gamma_6\gamma_6}
\eqno (16)
$$
is diagonal in $(\Gamma_3, \Gamma_6)$ and 
               $(\gamma_3, \gamma_6)$. Furthermore, 
the diagonal value of $X$ is independent of 
$\gamma_3$. More precisely, we have
$$
\eqalign{
  X =   \delta \left (\Gamma_6, \Gamma_3 \right ) \;
        \delta \left (\gamma_6, \gamma_3 \right ) \; 
&       (-1)^{2(k_3+k_6)} \; 
  \left ( [j_1] \; [j_4] \right )^{- 1/2} \; [\Gamma_3]^{-1} \; 
                                             [\Gamma_1] \cr 
& Z(j_1a_1, j_2a_2, k_3a_3, 
    j_4a_4, j_5a_5, k_6a_6; \Gamma_1 \Gamma_2 \Gamma_3), 
} 
\eqno (17)
$$
where 
$$
\eqalign{
Z(j_1a_1, j_2a_2, k_3a_3, & 
  j_4a_4, j_5a_5, k_6a_6; \Gamma_1 \Gamma_2 \Gamma_3) 
\; := \cr 
& \sum_\beta \; 
(j_2 a_2 \Gamma_2 + k_3 a_3 \Gamma_3 \vert j_1 a_1 \beta \Gamma_1 ) \ 
(j_5 a_5 \Gamma_2 + k_6 a_6 \Gamma_3 \vert j_4 a_4 \beta \Gamma_1 )^*
} 
\eqno (18)
$$
that reduces to the product of two isoscalar 
factors when $G^*$ is multiplicity free. 
}

It is also useful to express $Z$ and thus $X$ 
in terms of the $R_{\beta}$ symbols which are more symmetrical 
than the isoscalars factors. In this respect, by using (15) 
we can write (18) as 
$$
  \eqalign{
Z(j_1a_1, j_2a_2, k_3a_3, &
  j_4a_4, j_5a_5, k_6a_6; \Gamma_1 \Gamma_2 \Gamma_3) 
  = (-1)^{2(k_3+k_6)} \; [\Gamma_1] ^{-2} \; [j_1] \; [j_4] \cr 
  \sum_\beta \; \sum_{a_x \Gamma_x} \; \sum_{a_y \Gamma_y} \; 
& R
  \pmatrix{ 
  0            & j_1          & j_1 \cr
  \cr
  a_0 \Gamma_0 & a_1 \Gamma_1 & a_x \Gamma_x}^* \;
  R_\beta
  \pmatrix{ 
  k_3          & j_2          & j_1\cr
  \cr
  a_3 \Gamma_3 & a_2 \Gamma_2 & a_x \Gamma_x} \cr
& R
  \pmatrix{ 
  0            & j_4          & j_4\cr
  \cr
  a_0 \Gamma_0 & a_4 \Gamma_1 & a_y \Gamma_y}{\phantom {^*}} \;
  R_\beta 
  \pmatrix{ 
  k_6          & j_5          & j_4\cr
  \cr
  a_6 \Gamma_3 & a_5 \Gamma_2 & a_y \Gamma_y}^*
  }
  \eqno (19)
$$
that is in fact independent of the $U$ matrix of Equation (15). 

\avant 

\noindent {\bf 2. Intensity Formula}

\apres 

\noindent For a given transition operator $H_{\rm eff}$, 
we are now in a position to effectuate the summation over 
$\gamma$ and 
$\gamma'$ in (5) and, thus, to obtain a compact 
expression for the intensity $S_{i(\Gamma) \to f(\Gamma')}$. 
A very general expression for the effective operator $H_{\rm eff}$ 
is 
$$
H_{\rm eff} = 
\sum_{k_1 k_2 \cdots k_{p-1}} \, \sum_{t} \, \sum_{k_S k_L} \, \sum_{k} \, 
    C[k_1 k_2 \cdots k_{p-1} ; t ; k_S k_L ; k ] \, 
 \left( \{ {\bf P}^{(k_{p-1})} 
           {\bf X}^{(t      )} \}^{(k)} . \, 
           {\bf W}^{(k_Sk_L)k} \right),
\eqno (20)
$$
where $( \, . \, )$ is a scalar product involving electronic 
(${\bf W}^{(k_Sk_L)k}$) 
and nonelectronic (${\bf P}^{(k_{p-1})}$ and ${\bf X}^{(t)}$) tensors. 
In Equation (20), 
${\bf P}^{(k_{p-1})}$ denotes the polarization tensor defined by
$$
 {\bf P}^{(k_{p-1})} := \{ \cdots \{ \{     {\bf E}_1 
                                            {\bf E}_2 
                      \}^{(k_1    )}        {\bf E}_3
                      \}^{(k_2    )} \cdots {\bf E}_p 
                      \}^{(k_{p-1})} 
\eqno (21)
$$
that describes the coupling of the $p$ polarization 
vectors and that is entirely under the control of the 
experimentalist. The tensors ${\bf W}^{(k_Sk_L)k}$ and 
                             ${\bf X}^{(t      )}$ are 
relative to the ion and its environment: 
${\bf W}^{(k_Sk_L)k}$ is an electronic double tensor of 
spin    degree $k_S$, 
orbital degree $k_L$ and 
total   degree $k$
whereas 
${\bf X}^{(t      )}$ is a (single) tensor of the degree $t$, 
for instance, the ligand polarization tensor or the crystal-field 
tensor. (The tensor ${\bf X}^{(t      )}$ may result of the 
coupling of several $G$-invariant operators according to the 
order of the mechanism used.) 
The $C$ coefficients are 
model-dependent parameters that depend, among other things, on 
the initial, final and intermediate configurations as well as 
on the energies of the $p$ photons. 
Equation (20) unifies various models introduced in [1-4] for 
$p=1$ and in [5,~13-24] for $p=2$. It 
              can be applied to both (i) parity-allowed and 
                                    (ii) parity-forbidden transitions. 

(i) The particular case of intra-configurational two-photon 
transitions ($p=2$) is of special interest. 
For parity-allowed transitions within the 
configuration $n \ell^N$, it is sufficient 
to take $t = 0$ with ${\bf X}^{(0)} = 1$ 
in order to describe second-order             mechanisms 
when  $k_S=0$  and  second- plus third-order mechanisms 
when  $k_S=0$  and  $k_S \ne 0$. The situation where $t \ne 0$ 
permits to describe (other) third-order  mechanisms when $k_S   = 0$ 
                        and fourth-order mechanisms when $k_S \ne 0$. 
As a matter of fact, (20) gives 
back the model introduced by Axe [5] 
(with $t =   0$, $k_S =   0$) 
for second-order mechanisms and extended by different 
authors [14,~15,~17,~18,~20,~21] 
(with $t \ne 0$, $k_S \ne 0$) 
to take into account higher-order mechanisms. 

(ii) Two particular cases deserve a special attention, viz., 
the intra-configurational one-photon transitions and 
the inter-configurational two-photon transitions. For 
parity-forbidden transitions, ${\bf X}^{(t)}$ is either the ligand 
polarization tensor or the odd crystal-field tensor. 
In the case of one-photon transitions ($p=1$) within the 
configuration $n\ell^N$, (20) with $k_S = 0$ gives as a special case 
the operator implicitly considered by several people in 
order to explain intra-configurational one-photon transitions 
for $d^N$ [1,~2] or $f^N$ [3,~4] ions in crystals. 
In the case of inter-configurational two-photon transitions 
($p=2$) between the configuration $n \ell^N$ and the 
configuration $n \ell^{N-1} n' \ell'$ with 
$\ell + \ell'$ odd, (20) with $k_S=0$ 
yields as particular cases the 
operators introduced in connection with either a 
static  coupling mechanism [13,~16,~19,~22,~24] 
when ${\bf X}^{(t)}$ is the odd crystal-field   tensor or a 
dynamic coupling mechanism [19,~22] 
when ${\bf X}^{(t)}$ is the ligand polarization tensor. 

The transition matrix element for the operator (20) 
may be easily calculated owing to (11). We thus get 
$$
\eqalign{
M_{i(\Gamma \gamma) \to f (\Gamma' \gamma')} = & \sum_{\alpha' S' L' J' a'} \; 
                                                 \sum_{\alpha  S  L  J  a } 
\; (-1)^{J-J'}
\; c (\alpha' S' L' J' a' \Gamma' ; f)^* 
\; c (\alpha  S  L  J  a  \Gamma  ; i)\cr
& \sum_{k_1 k_2 \cdots k_{p-1}} \; 
  \sum_{t} \; \sum_{k_S k_L} \; \sum_{k} \; \sum_{a_0} \; 
C \left[ k_1 k_2 \cdots k_{p-1} ; t ; k_S k_L ; k \right] \; 
X^{(t)}_{a_0 \Gamma_0 \gamma_0} \cr 
&
[k]^{1 / 2} \; \left( \alpha' S' L' J' \Vert W^{(k_S k_L)k} \Vert 
                      \alpha  S  L  J  \right) \;
  \sum_{a'' \Gamma'' \gamma''} \; \sum_{b''}   \; 
{P}^{(k_{p-1})}_{a'' \Gamma'' \gamma''}    \cr 
& f
\pmatrix{
k                     & k_{p-1}               & t\cr
\cr
b'' \Gamma'' \gamma'' & a'' \Gamma'' \gamma'' & a_0 \Gamma_0 \gamma_0}^* \; 
f
\pmatrix{
J               & J'                 & k\cr
\cr
a \Gamma \gamma & a' \Gamma' \gamma' & b'' \Gamma'' \gamma''}^*, 
}
\eqno (22)
$$
where we have used the fact that the tensor ${\bf X}^{(t)}$ has 
only $G$-invariant components $X^{(t)}_{a_0 \Gamma_0 \gamma_0}$ 
distinguishable by the multiplicity label $a_0$. The right hand 
side of (22) can be expressed in terms of $\overline f$ 
coefficients in view of the connecting formula (arising from the 
comparison of (6) and (11)) between the $f$ and $\overline f$ symbols.

By introducing (22), in terms of $f$ or $\overline f$ 
symbols, into (5) and by using the corollary of 
Section 1, we obtain the intensity formula
$$
S_{i(\Gamma) \to f(\Gamma')} \; = \; 
                         \sum_{ \left\{ k   _i \right\} } 
                         \sum_{ \left\{ \ell_i \right\} } 
                         \sum_{r} \sum_{s} 
                         \sum_{\Gamma''} \; 
I[ \left\{ k   _i \right\} 
   \left\{ \ell_i \right\}
r s \Gamma'' ; \Gamma \Gamma'] \; 
\sum_{\gamma''} \; 
{P}^{(k   _{p-1})}     _{r        \Gamma'' \gamma''} \; \left( 
{P}^{(\ell_{p-1})}     _{s        \Gamma'' \gamma''}    \right)^* 
\eqno (23)
$$
(with $1 \le i \le p-1$), 
the form of which holds for both parity-allowed 
and parity-forbidden $p$-photon transitions. 
The intensity parameters $I$ in (23) are given by
$$
\eqalign{
I[ \left\{ k   _i \right\} 
   \left\{ \ell_i \right\}
r s \Gamma'' & ; \Gamma \Gamma'] \; := \; 
     \sum_{     J'      a'} \; 
     \sum_{     J       a } \; 
     \sum_{\bar J' \bar a'} \; 
     \sum_{\bar J  \bar a } \;
     \sum_{     k  \bar r } \;
     \sum_{\ell    \bar s } \cr 
& Y_k    (     J'      a' \Gamma' ,      J      a \Gamma, 
  \left\{    k_i \right\} r \bar r \Gamma'')   \ 
  Y_\ell (\bar J' \bar a' \Gamma' , \bar J \bar a \Gamma, 
  \left\{ \ell_i \right\} s \bar s \Gamma'')^* \cr
& Z(     J      a,      J'      a', k    \bar r, 
    \bar J \bar a, \bar J' \bar a', \ell \bar s ; \Gamma \Gamma' \Gamma''),
}
\eqno(24)
$$
where the $Y$ symbol is defined through 
$$
\eqalign{
Y_k    (     J'      a' \Gamma' ,    & J      a \Gamma, 
    \left\{    k_i \right\} r \bar r \Gamma'') 
\; := \;   \left( [J] [\Gamma''] \right)^{-1/2} \; 
          \left( [k] [\Gamma  ] \right)^{ 1/2} \; (-1)^{J - J'}
\sum_{\alpha' S' L'} \; 
\sum_{\alpha  S  L } \; \sum_{k_S k_L} \cr 
&  c (\alpha' S' L' J' a' \Gamma' ; f)^* \; 
   c (\alpha  S  L  J  a  \Gamma  ; i)   \; 
\left( \alpha' S' L' J' \Vert W^{(k_S k_L)k} \Vert 
       \alpha  S  L  J  \right)        \cr 
& \sum_{t a_0} \; 
C \left[ k_1 k_2 \cdots k_{p-1} ; t ; k_S k_L ; k \right] \; 
X^{(t)}_{a_0 \Gamma_0 \gamma_0} \; 
f
\pmatrix{
k                        & k_{p-1}             & t\cr
\cr
\bar r \Gamma'' \gamma'' & r \Gamma'' \gamma'' & a_0 \Gamma_0 \gamma_0}^* 
}
\eqno (25)
$$
which does not depend on $\gamma''$. 
The $I$ parameters in (24) can be calculated in an {\it ab initio} 
way or can be considered as phenomenological parameters. In both 
approaches, the following properties and rules are of central 
importance. 

\noindent PROPERTY 1. 
{\it In the general case, we have the (hermiticity) property
$$
I[\left\{ \ell_i \right\} \left\{ k_i \right\}sr\Gamma'';\Gamma \Gamma']\; = \; 
I[\left\{ k_i \right\} \left\{ \ell_i \right\}rs\Gamma'';\Gamma \Gamma']^* 
\eqno (26)
$$
that ensures that $S_{i(\Gamma) \to f(\Gamma')}$ is a real number.
}

\noindent PROPERTY 2. 
{\it In the case where the group $G$ is multiplicity free, 
we have the factorization formula
$$
            I[\left\{    k_i \right\} 
              \left\{ \ell_i \right\} r s \Gamma'' 
                                                   ; \Gamma \Gamma'] 
\; = \; \chi [\left\{    k_i \right\} r   \Gamma'' ; \Gamma \Gamma'] \; 
\>      \chi [\left\{ \ell_i \right\} s   \Gamma'' ; \Gamma \Gamma']^*, 
\eqno (27a)
$$
where the function $\chi$ is defined through 
$$
\chi [\left\{    k_i \right\} r \Gamma'' ; \Gamma \Gamma'] \; := \; 
\sum_{J'a'} \; 
\sum_{J a } \; \sum_{k \bar r} \ 
Y_k(J' a' \Gamma' , J a \Gamma, \left\{ k_i \right\} r \bar r \Gamma'') \ 
\, (J' a' \Gamma' + k \bar r \Gamma'' \vert J a \Gamma)
\eqno (27b)
$$
which follows from (24) and the factorized form of $Z$ (see (18)).
}

The number of independent parameters $I$ in the 
expansion (23) can be {\it a priori} determined 
from the two following selection rules 
used together with Properties 1 and 2.

\noindent RULE 1. 
{\it In order to have $S_{i(\Gamma) \to f(\Gamma')} \ne 0$, 
it is necessary that
$$
\Gamma'' \subset \Gamma'^* \otimes \Gamma \qquad \quad
\Gamma'' \subset (   k_{p-1})             \qquad \quad 
\Gamma'' \subset (\ell_{p-1})             \qquad \quad 
\Gamma'' \subset (   k      )             \qquad \quad 
\Gamma'' \subset (\ell      ), 
\eqno (28)
$$
where $(k_{p-1})$, $(\ell_{p-1})$, $(k)$ and $(\ell)$ are IRC's of the group 
$O(3)$ and $\Gamma'^*$ is the complex conjugate of
           $\Gamma'  $. 
}

\noindent RULE 2. 
{\it The sum over $ \left\{ k   _i \right\} $ 
              and $ \left\{ \ell_i \right\} $ in the intensity 
formula (23) is partially controlled by the following points. 
(i) The order of the mechanism used for describing the 
absorption processus: $k_{p-1}$ and $\ell_{p-1}$ cannot vanish 
if only $p$-order mechanisms, corresponding to $t=k_S=0$, are 
taken into consideration. (In this case, 
$k   _{p-1} = k   _L = k   $ and 
$\ell_{p-1} = \ell_L = \ell$.) Conversely, $k_{p-1}$ and $\ell_{p-1}$ 
may be zero if higher-order mechanisms, corresponding to 
$t   \ne 0$ and/or 
$k_S \ne 0$, are introduced. 
(ii) The nature of the photons: if the $p$ absorbed photons 
have the same polarization, then 
        $k_1    ( $and $\ell_ 1   ) = 0,2$ so 
that if $k_i    ( $or  $\ell_ i   ) = 0  $ we have 
        $k_{i+1}( $or  $\ell_{i+1}) = 1  $ and 
        $k_{i+2}( $or  $\ell_{i+2}) = 0,2$ 
when $i \ge 1$. 
}

\avant 

\noindent {\bf 3. Discussion and Closing Remarks} 

\apres

\noindent 
For low values of $k_{p-1}$ and $\ell_{p-1}$, 
there is no summation on $r$ and $s$, two 
branching multiplicity labels of type $a$, in 
the intensity formula (23). (The frequency of $\Gamma''$ in 
$(   k_{p-1})$ and 
$(\ell_{p-1})$ 
is rarely greater than 1 for $p \le 2$.) The group-theoretical 
selection rules (28) impose strong limitations 
on the summation over $\Gamma''$ in (23) once $\Gamma$ and $\Gamma'$ are
fixed and the range of values of 
$k$ and $\ell$ 
is chosen. 

The number of independent intensity parameters $I$ in 
(23) is determined by: 
(i) the number of photons 
and their nature (polarization, energy), cf.~Rules 1 and 2;
(ii) the symmetry group $G$, cf.~Rule 1; (iii) the symmetry property 
(26), 
cf.~Property 1; 
(iv) the use of $t  =   0$ and $k_S =   0$ ($p$-order    mechanism) or 
                $ t \ne 0$ and $k_S \ne 0$ (higher-order mechanisms), 
cf.~Rule 2; 
(v) the kind of the (weak-, intermediate- or 
strong-field) coupling used for the state vectors, cf.~(3). 
Points (i)-(ii) depend on external physical conditions. On the other hand,
points (iv) and (v) are model-dependent. In particular, in the case where the
$J$-mixing, cf.~point (v), can be neglected, a situation of interest for
lanthanide ions, the summations on $k$ and $\ell$ in (24) are reduced 
by the rule $|J-J'| \leq k, \ell \leq J+J'$, where $J$ and $J'$ are
the total angular momenta for the initial and final states,
respectively. Similar restrictions apply to $k_S$ and $k_L$ in 
(25) if the $S$- and $L$-mixing are neglected.

The computation, via Equations (18) or (19), (24) and (25), 
of the $I$ parameters is a difficult task in general. 
Therefore, they may be considered, at least in a
first step, as phenomenological parameters. In this respect, Equations 
(18) or (19), (24) and (25) 
should serve as a guide for reducing the number of $I$ parameters. 
Once the number of independent parameters $I$ in the intensity
formula (23) has been determined, we can obtain the polarization dependence 
of the intensity strength 
$S_{i(\Gamma) \to f(\Gamma')}$ by calculating the tensor products 
$P ^{(K)} _{a'' \Gamma'' \gamma''}$ 
(with $K = k_{p-1}, \ell_{p-1}$ and $a'' = r, s$) occurring in Equation 
(23). 

To close this Letter, some remarks are in order. In the 
particular case $p=1$ and $2$, the Hamiltonian model given 
by (20) and (21) unifies various models described in the 
literature 
for rare earth ions and 
transition-metal ions. The 
originality of this work rests on the 
use of symmetry adaptation methods for 
the chain $SU(2) \supset G^*$ in conjunction with a very general 
Hamiltonian model for describing simultaneous absorption of $p$ 
photons between Stark levels (rather than 
between $J$ levels). Nontrivial aspects of symmetry adaptation 
methods have been taken into account in a quantitative way: 
factorization {\it \`a la} Wigner-Eckart for the chain 
$SU(2) \supset G^*$, Racah's lemma for the corresponding 
coupling coefficients and orthogonality-completeness of the 
latter. As a net result, we have obtained the intensity formula 
(23) where the dynamics appears in the $I$ parameters and 
the geometry is contained in the $P$ factors describing the 
polarization dependence. 
The intensity formula (23) for multi-photon absorption can be 
extended to multi-photon emission as well as to Rayleigh and 
Raman scattering modulo some caution with the $C$ parameters in 
(25). The general form of (23) is also valid for other 
multi-photon processes, as for example the simultaneous 
absorption of several photons, certain by 
electric-dipole absorption and others by 
magnetic-dipole and/or 
electric-quadrupole absorption. 

A word should be said about previous works on this subject. 
The subject treated in the present Letter has been touched upon 
in [13,~21,~23-26] for $p=2$ and fully developed in the 
thesis by one of the authors (M.D.) for $p$ arbitrary. 
The accent has been put here on the intensity, rather than on 
the transition moment as in some works dealing with two-photon 
spectroscopy [13,~21]. Furthermore, this work is 
concerned with models and their use in connection with 
symmetry adaptation methods, rather than with qualitative 
symmetry considerations only. In this respect, our work 
represents a further important step besides the pioneer works by 
Inoue and Toyozawa [25], on one hand, and by Bader and Gold [26], on 
the other hand, where only symmetry considerations, arising from 
the group $G$ considered as an isolated entity, are introduced 
in the situation where $p=2$. 

\avant
\baselineskip = 0.59 true cm 
\parskip      = 0.11 true cm
\noindent {\bf References}

\apres

  \item{1.}
  Sugano, S., {\it Prog.~Theor.~Phys.~(Kyoto) Suppl.}~{\bf 14}, 66 (1960). 

  \item{2.}
  Griffith, J.S., {\it Mol.~Phys.}~{\bf 3}, 477 (1960). 

  \item{3.} 
  Judd, B.R., {\it Phys.~Rev.}~{\bf 127}, 750 (1962).

  \item{4.} 
  Ofelt, G.S., {\it J.~Chem.~Phys.}~{\bf 37}, 511 (1962). 

  \item{5.}
  Axe, J.D., Jr., {\it Phys.~Rev.}~{\bf 136}, A42 (1964).   

  \item{6.} 
  Flato,  M., {\it J.~Mol.~Spectr.}~{\bf 17}, 300     (1965).

  \item{7.} 
  Low, W.~and Rosengarten, G., 
            {\it J.~Mol.~Spectr.}~{\bf 12}, 319     (1964).

  \item{8.} 
  Kibler, M., {\it J.~Mol.~Spectr.}~{\bf 26}, 111     (1968); 
              {\it Int.~J.~Quantum Chem.}~{\bf 3}, 795     (1969). 

  \item{9.} 
  Kibler, M.~and Grenet, G., 
  {\it Int.~J.~Quantum Chem.}~{\bf 29}, 11      (1986);
                              {\bf 29}, 485     (1986). 

  \item{10.} 
  Kibler, M., {\it C.~R.~Acad.~Sc.~(Paris)} {\bf B 268}, 1221      (1969).

  \item{11.}
  Griffith, J.S., 
  {\it The Irreducible Tensor Method for Molecular Symmetry Groups}, 
  Prentice-Hall, Englewood Cliffs, New Jersey, 1962. 

  \item{12.} 
  Tang Au-chin and coll., 
  {\it Theoretical Method of the Ligand Field Theory}, 
  Science Press, Beijing, 1979. 

  \item{13.} 
  Apanasevich, P.A., Gintoft, R.I., Korolkov, V.S., 
  Makhanek, A.G.~and Skripko, G.A., 
  {\it Phys. Status Solidi (b)} {\bf 58 }, 745 (1973); 
  Makhanek, A.G., Korolkov, V.S.~and Yuguryan, L.A., 
  {\it Phys. Status Solidi (b)} {\bf 149}, 231 (1988).
 
  \item{14.} 
  Judd, B.R.~and Pooler, D.R., {\it J.~Phys.~C} {\bf 15}, 591 (1982).

  \item{15.}
  Downer, M.C.~and Bivas, A., {\it Phys.~Rev.~B} {\bf 28}, 3677 (1983).

  \item{16.} Gayen, S.K.,    Hamilton, D.S.~and Bartram, R.H., 
             {\it Phys.~Rev.~B} {\bf 34}, 7517 (1986). 

  \item{17.} 
  Reid, M.F.~and Richardson, F.S., 
  {\it Phys.~Rev.~B} {\bf 29}, 2830 (1984). 

  \item{18.}
  Sztucki, J.~and Str\c ek, W., {\it Phys.~Rev.~B} {\bf 34}, 3120 (1986).

  \item{19.} 
  Leavitt, R.C., {\it Phys.~Rev.~B} {\bf 35}, 9271 (1987). 

  \item{20.} 
  Smentek-Mielczarek, L.~and Hess, B.A., Jr., 
  {\it Phys.~Rev.~B} {\bf 36}, 1811 (1987). 

  \item{21.}
  Kibler, M.~and G\^acon, J.-C., {\it Croat.~Chem.~Acta} {\bf 62}, 783     
  (1989).

  \item{22.} 
  Sztucki, J.~and Str\c ek, W., 
  {\it Chem.~Phys.}~{\bf 143}, 347 (1990). 

  \item{23.} 
  Kibler, M.R., in {\it Symmetry and Structural Properties of Condensed Matter},
  W.~Florek, T.~Lulek and M.~Mucha, eds., 
  World Scientific, Singapore, 1991, p.~237. 

  \item{24.} 
  Daoud, M.~and Kibler, M., {\it J.~Alloys and Compounds} {\bf 193}, 219 
  (1993). 

  \item{25.} 
  Inoue, M.~and Toyozawa, Y., {\it J.~Phys.~Soc.~Japan} {\bf 20}, 363
  (1965). 

  \item{26.} 
  Bader, T.R.~and Gold, A., {\it Phys.~Rev.}~{\bf 171}, 997      (1968).

\bye